\documentclass[
    ,final            
  ]
  {aipproc}

\layoutstyle{6x9}

\begin{document}

\title{Study of nucleon resonances at EBAC@JLab}


\author{Hiroyuki Kamano}{
address={Excited Baryon Analysis Center (EBAC), Thomas Jefferson National Accelerator Facility,\\ Newport News, Virginia 23606, USA}
}

\begin{abstract}
We present the dynamical origin of the $P_{11}$ nucleon resonances
resulting from a dynamical coupled-channels (DCC) analysis
of meson production reactions off a nucleon target, which is conducted
at Excited Baryon Analysis Center (EBAC) of Jefferson Lab.
Two resonance poles are found in the energy region 
where the Roper resonance $P_{11}(1440)$ was identified.
Furthermore, the two resonance poles and the next higher resonance pole
corresponding to $P_{11}(1710)$ 
are found to originate from a single bare state.
\end{abstract}

\maketitle


\section{Introduction}

The excited nucleon states (collectively referred to as $N^\ast$)
are known to be a realization of 
the nonperturbative dynamics of the strong interaction.
An understanding of their spectrum and structure is a fundamental challenge
in the hadron physics.

The $N^*$ states, however, couple strongly to the meson-baryon continuum
states and appear only as resonance states in $\pi N$ and $\gamma N$ reactions.
Such a strong coupling to the continuum states
influence strongly the $N^*$ properties
and thus cannot be neglected in extracting information on the $N^\ast$ states 
from the data and giving physical interpretations.

To address this issue, an extensive research program of extracting 
the $N^\ast$ information through the comprehensive analysis
of $\pi N,~\gamma N,~N(e,e')$ reactions
is being conducted at Excited Baryon Analysis Center (EBAC) of Jefferson Lab.
The analysis is performed with
a dynamical coupled-channels (DCC) model developed in Ref.~\cite{msl07},
within which the couplings among relevant meson-baryon channels 
are fully taken into account.
The main objectives of EBAC are to extract spectrum of the $N^\ast$ states
and their form factors from the analysis
and provide reaction mechanism information necessary for interpreting 
the spectrum, structure, and dynamical origin of the $N^*$ states.
In this contribution we present our recent results on 
the dynamical origin of the $P_{11}$ nucleon resonances.

The EBAC-DCC analysis 
is based on a multi-channel and 
multi-resonance model~\cite{msl07}, within which 
the partial wave amplitudes of
$M(\vec p) + B(-\vec p) \to M'(\vec p') + B'(-\vec p')$ 
with $MB,M'B'=(\pi N,\eta N,\pi\Delta,\sigma N,\rho N)$
are calculated by the coupled-channels equations
(suppressing the angular momentum and isospin indices):
\begin{eqnarray}
T_{MB,M'B'}(p,p';E) &=& 
V_{MB,M'B'}(p,p';E)
\nonumber\\
&&
\!\!\!\!\!\!\!\!\!\!\!\!\!\!\!
\!\!\!\!\!\!\!\!\!\!\!\!\!\!\!
+\sum_{M''B''}\int dq q^2 V_{MB,M''B''}(p,q;E)
G_{M''B''}(q;E) T_{M''B'',M'B'}(q,p';E),
\label{eq:lseq}
\end{eqnarray}
where $G_{MB}(q;E)$ is the Green function of the $MB$ channel,
expressed as 
$G_{MB}(q;E)=1/[E-E_M(q)-E_B(q) + i\epsilon]$
for the stable $\pi N$ and $\eta N$ channels and
$G_{MB}(q;E)=1/[E-E_M(q)-E_B(q) - \Sigma_{MB}(q;E)]$
for the unstable $\pi\Delta$, $\rho N$, and $\sigma N$ channels.
The imaginary part of the self-energy $\Sigma_{MB}(q,E)$ 
comes from the $\pi\pi N$ unitarity cut.

The $MB\to M'B'$ transition potential is defined by
\begin{equation}
V_{MB,M'B'}(p,p';E) = 
v_{MB,M'B'}(p,p') + 
\sum_{N^\ast_i}
\frac{\Gamma^\dag_{N^\ast_i,MB}(p)\Gamma_{N^\ast_i,M'B'}(p')}{E-m^0_{N^\ast_i}},
\label{eq:pot}
\end{equation}
where $m^0_{N^\ast_i}$ and $\Gamma^\dag_{N^\ast_i,MB}(p)$
represent the mass and $N^\ast_i \to MB$ decay vertex
of the $i$-th bare $N^\ast$ state in a given partial wave, respectively.
The first term $v_{MB,M'B'}(p,p')$ is 
the (energy-independent) meson-exchange potentials,
which are derived from the effective Lagrangians
making use of the unitary transformation method~\cite{msl07,sl09};
the second term describes $MB\to M'B'$ transitions 
through the bare $N^\ast$ state, $MB\to N^\ast\to M'B'$.

The $MB\to M'B'$ amplitude (\ref{eq:lseq}) is a basic ingredient to construct
all single and double meson production reactions
with the initial $\pi N$, $\gamma N$, $N(e,e')$ states.
The hadronic and electromagnetic parameters of our current model 
have been fixed by the $\pi N$ scattering~\cite{jlms07} up to $W=2$ GeV
and $\gamma p\to\pi N$~\cite{jlmss08} and
$ep\to e'\pi N$~\cite{jklmss09} up to $W=1.6$ GeV, respectively, and 
the model has been applied to $\pi N\to \pi\pi N$~\cite{kjlms09} and 
$\gamma N\to\pi\pi N$~\cite{kjlms09-2}.

The resonance pole positions can be obtained as
zeros of the determinant of the inverse of 
the dressed $N^\ast$ propagator:
\begin{equation}
[D^{-1}(E)]_{i,j} = (E - m^0_{N^*_i})\delta_{i,j} - [M(E)]_{i,j},
\label{eq:nstar-selfe}
\end{equation}
where the self-energy of the dressed $N^\ast$ state given by
\begin{equation}
[M(E)]_{i,j}=
\sum_{MB}
\int q^2 dq 
\bar{\Gamma}_{N^*_j \to M B}(q;E) G_{MB}(q,E) {\Gamma}_{MB \to N^*_i}(q),
\label{eq:nstar-g}
\end{equation}
with $\bar\Gamma_{N^\ast\to MB}$ being the dressed $N^\ast\to MB$ vertex
defined in Ref.~\cite{msl07}.
To search for zeros of $\det[D^{-1}(E)]$ for complex $E$,
we need to make an analytic
continuation of the amplitudes.
The analytic continuation method we used
is described in detail in Refs.~\cite{ssl09,ssl09-2} and
will not be discussed here.
The pole positions of all nucleon resonances 
below $W=2$ GeV extracted from the current EBAC-DCC analysis 
can be found in Ref.~\cite{sjklms09}.

\section{$P_{11}$ nucleon resonances from EBAC-DCC analysis}

\subsection{Two-pole structure of the Roper resonance}

\begin{table}[b]
\begin{tabular}{ccc}
\hline
\tablehead{1}{c}{b}{}&
\tablehead{1}{c}{b}{Pole position\\(MeV)} &
\tablehead{1}{c}{b}{Location on the complex-$E$ plane\\
($\pi N,\eta N,\pi\pi N,\pi\Delta,\rho N,\sigma N$)
\tablenote{p = physical-sheet, u = unphysical-sheet}}
\\
\hline
A& $1357-i76$  & (u,p,u,u,p,p)\\
B& $1364-i105$ & (u,p,u,p,p,p)\\
C& $1820-i248$ & (u,u,u,u,u,p)\\
\hline
\end{tabular}
\caption{$P_{11}$ resonance poles below $W=2$ GeV from the EBAC-DCC analysis.}
\label{tab:p11-poles}
\end{table}

\begin{figure}[t]
\includegraphics[width=.54\textwidth]{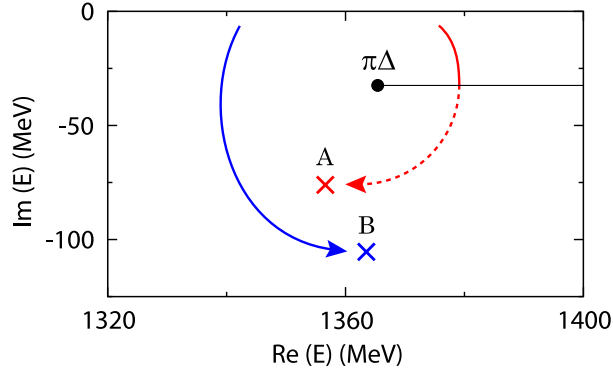}
\caption{
Two-pole structure of the Roper resonance.
The filled circle is the $\pi\Delta$ branch points;
the $\pi\Delta$ cut (solid line) runs in parallel 
with the real energy axis.
The poles A and B are located on the unphysical and physical sheets
with respect to the $\pi\Delta$ channel, respectively.
}
\label{fig:roper}
\end{figure}

The $P_{11}$ partial wave of the $\pi N$ scattering has been investigated 
with a particular interest in the literatures. 
This is mainly due to the controversy over the
dynamical origin of the mysterious Roper resonance $N^\ast(1440)$.
In the energy region below 2 GeV, 
we found three $P_{11}$ resonance poles relevant to the observables
as listed in Table~\ref{tab:p11-poles}.
Furthermore, two of the three resonance poles are found to be near 
the Roper resonance energy,
$E_A=1357-i76$ MeV and $E_B=1364-i105$ MeV, which are indicated as
points A and B in Fig.~\ref{fig:roper}, respectively.

At a first glance the reader might think the two poles are very close 
to each other because $|E_A-E_B|\sim 30$ MeV. 
However, in reality they locate in the different Riemann sheets with respect to 
the $\pi\Delta$ branch point and thus are ``far'' from each other:
The pole A locates in the sheet continued analytically from the 
upper side of the $\pi\Delta$ cut (unphysical sheet), whereas the pole B locates
in the sheet directly connected with the physical real energy axis 
(physical sheet).
(We take branch cuts to be in parallel with the positive direction
of the real energy axis as shown in Fig.~\ref{fig:roper} for the $\pi\Delta$
channel.)

The two Roper poles will not be observed as clear peaks
because of the analytic structure of the complex energy
plane induced by the $\pi\Delta$ branch point.
Nevertheless, those poles are still close to the physical region and 
expected to have a non-negligible contribution to the observables.
To prove this we need to make a detailed examination of 
the decay vertices into $\pi\pi N~(=\pi\Delta,\rho N,\sigma N)$ 
as well as $\pi N$.
This is under investigation and will be presented elsewhere.

It is noted that several groups have also reported 
similar two-pole structures of the Roper 
resonance~\cite{cmb,said,juelich-npa}.

\subsection{Dynamical origin of $P_{11}$ nucleon resonances}

Another important finding from our analysis is 
about the dynamical origin of the $P_{11}$ nucleon resonances. We find 
that the two Roper poles (A and B in Table.~\ref{tab:p11-poles})
and the next higher resonance pole corresponding to $N^\ast(1710)$
(C in Table~\ref{tab:p11-poles}), are generated from 
a single bare state as a result of its coupling to 
the meson-baryon continuum states.
Theoretically, Eden and Taylor already pointed out four decades ago
that multi-channel reactions can generate many resonance poles
from a single bare state~\cite{eden}.
In most cases, only one of the poles appears close to the physical region.
However, depending on a given reaction dynamics,
more than one pole can appear to have a physical significance.
Just few of such evidences were reported in the past
(see e.g., Ref.~\cite{morgan}).
Our result suggests that the $P_{11}$ nucleon resonances may be an
important addition.

To examine how the three $P_{11}$ poles evolve from a single bare state 
dynamically, we trace the zeros of 
$\det[\hat D^{-1}(E)]=\det[E-m_{N^\ast}^0-\sum_{MB}y_{MB}M_{MB}(E)]$
in the region $0\leq y_{MB}\leq 1$,
where $M_{MB}(E)$ is the $MB$-loop contribution to
the $N^\ast$ self-energy $M(E)$ defined in Eq.~(\ref{eq:nstar-g}).
Each $y_{MB}$ is varied independently to find continuous evolution
paths through the various Riemann sheet on which the analytic 
continuation method is valid.

\begin{figure}[t]
\includegraphics[width=.54\textwidth]{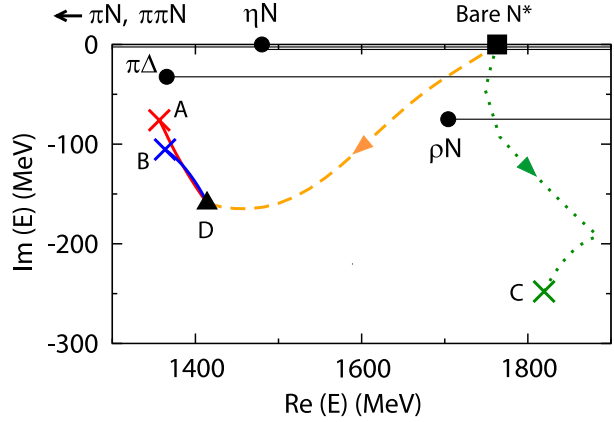}
\caption{
Pole trajectory of the $P_{11}$ nucleon resonances below $W=2$ GeV.
See the text for the description of the figure.
}
\label{fig:trajectory}
\end{figure}

By setting all $y_{MB}$'s to slightly positive from zero, 
the bare state (the filled square at $E=1763$ MeV in Fig.~\ref{fig:trajectory}) 
couples to all $MB$ channels and many poles are generated
according to the discussion by Eden and Taylor~\cite{eden}.
One of them appears on the $\eta N$-unphysical, 
$\rho N$-unphysical, and $\pi\Delta$-unphysical sheet
and it moves to the pole C 
by further varying all $y_{MB}$ to one 
(the dotted curve in Fig.~\ref{fig:trajectory}).

Similarly we can trace how the two Roper poles evolve from the
same bare state.
It is instructive to see this by first keeping $y_{\pi\Delta}$ zero
and varying the other $y_{MB}$'s from zero to one,
which means that the coupling to the $\pi\Delta$ is off in the variation.
With this variation we can trace another pole trajectory moving on
the $\eta N$-physical and $\rho N$-physical sheet 
along the dashed curve in Fig.~\ref{fig:trajectory}
from the bare position to the point D with Re$(E_D)\sim 1400$ MeV
(the filled triangle in Fig.~\ref{fig:trajectory}).
It is noted that the poles on the $\eta N$-physical sheet 
are far from the physical real energy axis above the $\eta N$ threshold,
while those are the nearest below the threshold.
Therefore this another pole on the $\eta N$-physical sheet, 
moving from the bare point to the point D along the dashed curve,
becomes the nearest resonance pole as a result of crossing the 
$\eta N$ threshold.
By further varying $y_{\pi\Delta}:0\to 1$, the trajectory splits into 
two trajectories: One moves to the pole A on the $\pi\Delta$ unphysical 
sheet and the other to the pole B on the $\pi\Delta$ physical sheet.
This indicates that the coupling to the $\pi\Delta$ channel is 
essential for the two-pole structure of the Roper resonance.
In this way, we observe that all the three $P_{11}$ resonance poles 
are connected to the same bare $N^\ast$ state at $E=1763$ MeV.

\section{summary}
We have investigated the dynamical origin of 
the $P_{11}$ nucleon resonances 
extracted recently from the EBAC-DCC analysis.
Our main findings are:
1) the Roper resonance is associated with the two resonance poles,
and
2) the two Roper resonance poles and the next higher resonance pole
corresponding to $N^\ast(1710)$ originate from a single bare state,
indicating that in this case the naive one-to-one correspondence between 
the bare states and the observed resonance poles is not applied.
We have demonstrated the critical role played by 
the non-trivial multi-channel reaction mechanisms 
in interpreting the dynamical origin of nucleon resonances.
Our results have provided new insights in understanding
the spectrum of the $N^\ast$ states.

\begin{theacknowledgments}
The author would like to thank B.~Juli\'a-D\'{\i}az, T.-S.~H.~Lee, 
A.~Matsuyama, T.~Sato, and N.~Suzuki for their collaborations at EBAC.
This work was supported by 
the U.S. Department of Energy, Office of Nuclear Physics Division, under 
Contract No. DE-AC05-06OR23177 
under which Jefferson Science Associates operates the Jefferson Lab.
\end{theacknowledgments}

\bibliographystyle{aipproc}   

\end{document}